\documentclass[%
 reprint,
 superscriptaddress,
 showpacs,preprintnumbers,
 nofootinbib,
 amsmath,amssymb,
 aps,
]{revtex4-1}

\usepackage{graphicx}
\usepackage{dcolumn}
\usepackage{bm}
\usepackage{color}

\begin{document}


\title{Future experimental improvement for the search of \\ lepton-number-violating processes in the $e\mu$ sector}


\author{Beomki Yeo}
 \email{byeo@kaist.ac.kr}

\affiliation{Department of Physics, Korea Advanced Institute of Science and Technology (KAIST), Daejeon 34141, Republic of Korea} 
  
\author{Yoshitaka Kuno}
 \email{kuno@phys.sci.osaka-u.ac.jp}

\affiliation{Department of Physics, Graduate School of Science, Osaka University,  Toyonaka, Osaka 560-0043, Japan}

\author{MyeongJae Lee}
\email{myeongjaelee@ibs.re.kr}

\affiliation{Center for Axion and Precision Physics Research, Institute for Basic Science (IBS), Daejeon 34051, Republic of Korea}
 
\author{Kai Zuber}
\email{zuber@physik.tu-dresden.de}  

\affiliation{Institute for Nuclear and Particle Physics, Technische Universit{\"a}t Dresden, Germany}

\date{\today}

\begin{abstract}
The conservation of lepton flavor and total lepton number are no longer guaranteed in the Standard Model after the discovery of neutrino oscillations. $\mu^- + N(A,Z) \rightarrow e^+ + N(A,Z-2)$ conversion in a muonic atom is one of the most promising channels to investigate the lepton number violation processes, and measurement of the $\mu^--e^+$ conversion is planned in future  $\mu^--e^-$ conversion experiments with a muonic atom in a muon-stopping target. This letter discusses experimental strategies to maximize the sensitivity of the $\mu^--e^+$ conversion experiment by introducing the new requirement of the mass relation of $M(A,Z-2)<M(A,Z-1)$, where $M(A,Z)$ is the mass of the muon-stopping target nucleus, to eliminate the backgrounds from radiative muon capture. The sensitivity of the $\mu^--e^+$ conversion is expected to be improved by four orders of magnitude in forthcoming experiments using a proper target nucleus that satisfies the mass relation. The most promising isotopes found are $^{40}$Ca and $^{32}$S.
\end{abstract}

\pacs{13.35.Bv, 14.60.Ef, 36.10.Ee}
\maketitle



\section{Introduction}
Since lepton flavor violation was confirmed by the discovery of neutrino oscillation, interest has considerably shifted to 
the whole leptonic sector in terms of the search for new physics beyond the Standard Model (SM). Anomalies in the leptonic sector governed by new physics have been studied within three major phenomena: (1) lepton universality violation (LUV), (2) charged lepton flavor violation (CLFV), (3) and lepton number violation (LNV). The SM, which preserves the lepton universality, predicts that three generations of leptons behave consistently within the electroweak interaction. However, recent measurements of $\bar{B}\rightarrow D^{(*)}l^- \bar{\nu_l}$ \cite{BaBar2013,Belle2014,Belle2016,LHCb2015} and $B^+\rightarrow K^+ l^+ l^-$ \cite{LHCbKaon2014} have shown non-trivial discrepancies ($4  \sigma$ and $2.6 \sigma$, respectively) to the SM predictions, showing the possibility of LUV in new physics \cite{Nature2017}. 
 An interesting implication of LUV is that experimentally observable CLFV phenomena may emerge from new physics \cite{Fajfer2012,Glashow2015}. Although the processes of CLFV can occur by neutrino mixing in the SM, it should be noted that the rates of the SM contributions were found to be extremely small, on the order of $\mathcal{O}(10^{-54})$ because of small neutrino masses. Therefore, CLFV processes have been investigated through the various muon decay channels: $\mu^{-}-e^{-}$ conversion, $\mu^{+}\rightarrow e^{+}+\gamma$ decay and $\mu^{+} \rightarrow e^{+}+e^{+}+e^{-}$ decay in the expectation of a discovery of new physics \cite{Kuno2001}. 
\\
\indent The observation of LNV would provide crucial evidences on the small neutrino mass ($\lesssim$ eV). The LNV processes, with the change of lepton number by two units $(\Delta L=2)$, can be mediated by Majorana neutrinos through type-1 seesaw mechanism or new particles appearing at a high energy scale ($>$ TeV). These phenomena have been explored mostly through $0\nu\beta\beta$ decay \cite{DellOro2015}, which corresponds to the LNV process in the $ee$ sector. LNV processes in other sectors also have been searched with muon-to-positron conversion $\mu^- + N(A,Z) \rightarrow e^+ + N(A,Z-2)$ \cite{Copper1972,  Sulfur1978, Iodine1980, Sulfur1982, Titanium1988, Titanium1993, Titanium1998} and rare Kaon decays such as $K^+\rightarrow \mu^+ \mu^+ \pi^-$ \cite{RareKaon2000,RareKaon2011,RareKaon2017,Zuber2000}, while their experimental limits are far behind that of $0\nu\beta\beta$ decay, as shown in Table. \ref{tab:LNV_upperlimit}. 
\begin{table}[]
\centering
\caption{The experimental limits (as 90$\%$ C.L.) of some selected LNV processes. $(*)$ denotes the excited states of the daughter nucleus. \newline}
\begin{ruledtabular}
\begin{tabular}{llc}
       Process            &  Experimental limit    & Ref     \\ \hline
        $0\nu\beta\beta \textrm{ } (^{76}\textrm{Ge})$ & $\textrm{Half-life} > 5.3 \times 10^{25} \textrm{ yr}$  & \cite{Gerda2017} \\
       $0\nu\beta\beta$ $\textrm{ } (^{136}\textrm{Xe})  $ & $\textrm{Half-life} > 1.07 \times 10^{26} \textrm{ yr}$  & \cite{KamlandXen2016} \\
        $\mu^-+\textrm{Ti}\rightarrow e^++ \textrm{Ca}$      & $Br< 1.7\times10^{-12}$ & \cite{Titanium1998} \\ 
        $\mu^-+\textrm{Ti}\rightarrow e^++\textrm{Ca}^{*}$  &  $Br< 3.6\times10^{-11}$ & \cite{Titanium1998} \\
       $K^+\rightarrow \mu^+ \mu^+ \pi^-$ & $Br< 8.6 \times 10^{-11}$ & \cite{RareKaon2017}       \\
\end{tabular}
\end{ruledtabular}
\label{tab:LNV_upperlimit}
\end{table}
\\
\indent Nevertheless, the $\mu^--e^+$ conversion is worth investigating further for two reasons: (1) The $\mu^--e^+$ conversion is discoverable if  the LNV process is more likely to occur in flavor off-diagonal sectors, e.g., $e\mu$ sector, as implied by recent studies \cite{Berryman2016,Geib2017, Geib2017_2}. Several theories beyond the SM of particle physics, such as the Majorana neutrino, the doubly charged singlet scalar model \cite{Chen2007, King2014}, and the left-right symmetric model \cite{Pritimita2016} have been suggested as feasible theories for the $\mu^--e^+$ conversion. (2) In principle, the experimental sensitivity of the $\mu^--e^+$ conversion can significantly increase with the future $\mu^--e^-$ conversion experiments because the event signatures $(e^\pm)$ of both physics processes can be easily distinguished by the charge identification. 
However, in the upcoming COMET and Mu2e experiments\footnote{Both the COMET Phase-1 and the Mu2e experiments can measure the $\mu^--e^+$ and $\mu^--e^-$ conversion simultaneously, while the COMET Phase-2 experiment may need to change the polarity of the dipole magnetic field in the detector solenoid.} \cite{COMET_CDR_2009, COMET_TDR_2015, Mu2e_TDR_2014}, which were originally designed to search for the $\mu^--e^-$ conversion with the sensitivity of $\mathcal{O}(10^{-16})$, a similar scale of the sensitivity for the $\mu^--e^+$ conversion can only be achievable in the case of employing a proper material for a muon-stopping target nucleus. This limitation of $\mu^--e^+$ conversion sensitivity is due to the backgrounds from radiative muon capture (RMC). 
For example, as we will show later, the sensitivity improvement is less than a factor of ten in the case of an aluminum stopping target, which is the baseline design of the COMET and Mu2e experiments. In this study, this kind of limitation was surmounted by selecting a proper muon-stopping target nucleus, which suppresses the RMC background and improves the sensitivity by four orders of magnitude over the current limit. 
\\
\indent 
Not only the particle physics models of $\mu^--e^+$ conversion but also effects from nuclear physics should be carefully considered because the atomic number of the final state nucleus changes by two units after the conversion. This means that the $\mu^--e^+$ conversion is associated with nuclear interaction, which is represented as a nuclear matrix element in the calculation of the decay rate. Another aspect of nuclear interaction is that the nucleus in the final state after the $\mu^--e^+$ conversion is divided into two cases in which a daughter nucleus stays in the ground state or enters excited states. The transition to the ground state of the daughter nucleus may not be dominating since the coherence of the nucleus, which in the $\mu^--e^-$ conversion, is not expected in the $\mu^--e^+$ conversion. Nevertheless, in this work, we focus on the case of the ground state since the excited states suffer more from  background influence, and the momentum spectrum of signals for excited states is not understood well due to the uncertainty of nuclear physics. 
\\
\indent The remainder of this letter is organized as follows: In Sec. \ref{sec:target_nucleus}, we introduce the new requirement for the muon-stopping target nucleus with the mass of $M(A,Z)$ to suppress the background and investigate candidates of the target nucleus that meet this requirement. In Sec. \ref{sec:sensitivity}, the experimental sensitivities (the average upper limits) are estimated and compared among the target nucleus candidates with simulation results. A summary follows in Sec. \ref{sec:summary}.

\section{\label{sec:target_nucleus}Target Nucleus Candidates for the $\mu^--e^+$ conversion}
\subsection{Mass relation of target nucleus}
The principle of CLFV experiments based on a pulsed muon beam are as follows. A pulsed proton beam hits a pion production target to generate a bunch of pions. The negative pions are captured by a solenoidal magnetic field and are sent to a muon-stopping target, while most of the pions decay into muons during transport. These muons are stopped at the muon-stopping target, forming a muonic atom, and subsequently cascade down to the 1s ground state, followed either by muon decays in the 1s orbit of a muonic atom (decay-in-orbit=DIO), or by muon captures by a nucleus. A single positron emission from the $\mu^--e^+$ conversion is the signature of the process.
The positron is measured during the delayed time interval of the bunch period to avoid huge backgrounds by the primary beam during the prompt interval of the bunch period. For the transition to the ground state of the daughter nucleus, the signal positron is mono-energetic, and its energy $(E_{\mu^-e^+})$ is given by 
\begin{equation}\label{eq:sig_kinematics}
E_{\mu^-e^+} = m_\mu + M(A,Z)-M(A,Z-2) -B_\mu - E_{recoil}\, ,
\end{equation}
where $m_\mu$, $B_\mu$, and $E_{recoil}$ are the muon mass, the 1s binding energy of the muonic atom, and the recoil energy of the nucleus, respectively. Here, $M(A,Z)$ is the mass of the target nucleus, and $M(A,Z-2)$ is the mass of the ground state of the daughter nucleus. In the case of the $\mu^--e^-$ conversion, the signal energy is $E_{\mu^-e^-} = m_\mu -B_\mu - E_{recoil}$, where the mass terms in Eq. (\ref{eq:sig_kinematics}) are canceled out since the  target and the daughter nuclei are the same.
\\
\indent There are two major sources of background in the $\mu^--e^\pm$ detection: (1) DIO, and (2) RMC, as indicated by Ref. \cite{Copper1972,Sulfur1978, Sulfur1982, Titanium1988, Titanium1993, Titanium1998}. DIO has an endpoint energy $(E^{end}_{DIO})$, the same as $E_{\mu^-e^-}$, and can emit a high energy $e^-$ near the endpoint energy, which is an intrinsic background for the $\mu^--e^-$ detection. It can also fake the signal in the $\mu^--e^+$ detection 
when the charge is misidentified. However, it is expected that charge misidentification rarely occurs because of the high 
resolution of the tracking detectors. In the case of RMC, the emitted $\gamma$ can generate a high energy $e^-$ or $e^+$ after an assymmetric  pair production being an intrinsic background for both of the $\mu^--e^\pm$ conversion. Its endpoint energy $(E^{end}_{RMC})$ is kinematically given by
\begin{equation}\label{eq:rmc_kinematics}
E^{end}_{RMC} = m_\mu + M(A,Z) - M(A,Z-1) - B_\mu - E_{recoil}\, ,
\end{equation}
where $M(A,Z-1)$ is the mass of the daughter nucleus of RMC. The background from RMC becomes negligible when the corresponding signal energy ($E_{\mu^-e^\pm}$) is higher than $E^{end}_{RMC}$. Since a simultaneous search for both conversions is desired, two mass relations between nuclei are required to avoid the RMC background: (1) $M(A,Z-2) < M(A,Z-1)$ for the $\mu^--e^+$ conversion, and (2) $M(A,Z) < M(A,Z-1)$ for the $\mu^--e^-$ conversion. The latter requirement is generally satisfied for most of the stable nuclei, but the number of nuclei satisfying the former is limited because the daughter nucleus of the $\mu^--e^+$ conversion is usually less stable than that of RMC. However, this tendency can be reversed when even-even nuclei are used as the target material since the nucleons in the daughter nucleus of the $\mu^--e^+$ conversion, which is an even-even nucleus again, can bind more tightly due to the nuclear pairing force, whereas this is not the case for RMC with the odd-odd daughter nucleus. This consideration is similar to the target selection in the $0\nu\beta\beta$ decay experiments which require the mass relations of  $M(A,Z)>M(A,Z+2)$ and $M(A,Z)<M(A,Z+1)$ to enable the double beta decay, and forbid the single beta decay, respectively. 

\subsection{\label{ssec:target_material}Search for the target nucleus candidates}
\begin{table}[t]
\centering
\caption{Stopping-target nucleus candidates whose $E_{\mu^-e^+}$ is higher than, or comparable to, $E^{end}_{RMC}$. If more than two isotopes satisfy the criteria, only one isotope with the highest natural abundance (N.A.) is listed. Nuclear masses required for the calculations are referred from AME2016 data \cite{AME2016}. Aluminum, which is the counterexample, is listed because it is considered the muon-stopping target nucleus in the upcoming CLFV experiments. \newline}
\begin{ruledtabular}
\begin{tabular}{cccccccc}
Atom      & $E_{\mu^-e^+}$ &  $E_{\mu^-e^-}$ & $E^{end}_{RMC}$  & N.A.  & $f_{cap}$ & $\tau_{\mu^-}$ & $A_T$ \\
  & (MeV) & (MeV) & (MeV) & ($\%$) & ($\%$) & (ns)  \\ \hline
$^{27}\textrm{Al}$   & 92.30  & 104.97   & 101.34 & 100 & 61.0 & 864 & 0.191 \\  
$^{32}\textrm{S}$    & 101.80 & 104.76   & 102.03  & 95.0 & 75.0 & 555 & 0.142  \\ 
$^{40}\textrm{Ca}$   & 103.55 & 104.39   & 102.06  & 96.9 & 85.1 & 333 & 0.078 \\
$^{48}\textrm{Ti}$   & 98.89  & 104.18   & 99.17  & 73.7 & 85.3 & 329 & 0.076 \\
$^{50}\textrm{Cr}$   & 104.06 & 103.92   & 101.86  & 4.4  & 89.4 & 234 & 0.038 \\
$^{54}\textrm{Fe}$   & 103.30 & 103.65   & 101.93  & 5.9 & 90.9   & 206   & 0.027 \\
$^{58}\textrm{Ni}$   & 104.25 & 103.36   & 101.95 & 68.1 & 93.1 & 152 & 0.009\\
$^{64}\textrm{Zn}$   & 103.10 & 103.04   & 101.43 & 48.3 & 93.0 & 159 & 0.011 \\
$^{70}\textrm{Ge}$   & 100.67 & 102.70   & 100.02  & 20.8  & 92.7 & 167 & 0.013 \\ 
\end{tabular}
\end{ruledtabular}
\label{tab:target_candidate}
\end{table}
\indent  Table \ref{tab:target_candidate} lists the candidate target nuclei with atomic mass $\leq 70$ that satisfy the requirements. Heavier nuclei were not considered due to their shorter 
lifetimes of muonic atoms, leading to lower efficiencies in the finite time window of measurements, as explained in the next paragraph. In the present calculation of each energy value, $B_\mu$ was obtained by assuming a point-like nucleus while this may not hold for heavier nuclei due to the larger size of the nucleus, and further corrections are required \cite{Borie1982}. In Table \ref{tab:target_candidate}, $E^{end}_{RMC}$ from Eq. (\ref{eq:rmc_kinematics}) assumes RMC without an additional nucleon emission. RMC with nucleon emission can also generate backgrounds if its endpoint energy is higher than $E_{\mu^-e^+}$ or $E_{\mu^-e^-}$. However, this process does not generate additional backgrounds in most cases because the binding energy per nucleon is around 7--9 MeV for the stable nuclei, which means that the endpoint energy is lowered by a similar amount.
\\
\indent 
There are other requirements from an experimental point of view. For example, the muon capture rate $(f_{cap})$ and the muonic-atom lifetime $(\tau_{\mu^-})$ of each nucleus listed \cite{Measday2001, Suzuki1987} in Table \ref{tab:target_candidate} should be taken into account because $f_{cap}$ is proportional to the number of signal events, and $\tau_{\mu^-}$ is an important factor to determine the event acceptance in the time window of measurement $(A_T)$. The values of $A_T$ in Table \ref{tab:target_candidate} were calculated with a mathematical toy model with following assumptions: the bunch period $(t_B)$ of the muon beam of 1 $\mu$s, the timing window $([t_1,t_2])$ from 700 ns to 1 $\mu$s, and the uniform time distribution of muons with the bunch size of 100 ns. Then, $A_T$ is $N_{\textrm{time}}/N_{\textrm{total}}$, where $N_{\textrm{total}}$ is the number of stopped muons in the target with the single muon bunch, and $N_{\textrm{time}}$ is the number of decaying muons during the timing window. $N_{\textrm{time}}$ is given by $\sum_{n=1}^{\infty} \int^{t2+t_B(n-1)}_{t_1+t_B(n-1)} N(t)dt$, where $N(t)$ is the time distribution of exponential decays of muons  convoluted by the uniform time distribution of muons.
\\
\indent
Natural abundance is another important characteristic in the target selection for two reasons. First, the background from other isotopes can contaminate the signal. Second, the signal itself can be dispersed into a broader spectrum unless the natural abundance of the candidate isotope is high enough. Considering these requirements, $^{32}\textrm{S}$ and $^{40}\textrm{Ca}$ may be the most promising candidates because of their relatively high natural abundances and $A_T$, while the other candidate isotopes still can be considered by appropriate 
enrichment techniques.

\section{\label{sec:sensitivity}Experimental Sensitivities of Target Nucleus Candidates}\label{sec:bg_estimation}
In this section, the experimental sensitivities of target nucleus candidates are estimated assuming that the positron events only occur by the $\mu^--e^+$ conversion and RMC. The number of accepted positrons from the $\mu^--e^+$ conversions $(N_{\mu^-e^+})$ can be estimated by 
\begin{equation}\label{eq:N_SIG}
N_{\mu^-e^+} \sim  N_{\mu^-\textrm{stop}} \times f_{cap} \times Br(\mu^--e^+) \times \mathcal{E},
\end{equation}
where $N_{\mu^-\textrm{stop}}$ is the total number of the stopped muons in the target, $Br(\mu^--e^+)$ is the branching ratio of the $\mu^--e^+$ conversion, in which daughter nucleus stays in the ground state, and $\mathcal{E}$ is the net acceptance of signal positrons in the detector. $\mathcal{E}$ is assumed to be the same for the $\mu^--e^+$ conversion and RMC positrons. 
\\
\indent
The energy spectrum of RMC photons can be represented by \cite{Primakoff1959}
\begin{equation}\label{eq:RMC_Primakoff}
P(x) \simeq C(1-2x+2x^2)x(1-x)^2, \quad x=\frac{E_\gamma}{E^{end}_\gamma},
\end{equation}
where $C$ is the normalization constant determined from the results of previous experiments \cite{Dobeli1988,Armstrong1992,Bergbusch1999}, and $E^{end}_{\gamma}$ is the endpoint energy of RMC photons. In each experiment, the experimental values of $E^{end}_{\gamma}$ were obtained by fitting the photon energy distribution with the shape of Eq. (\ref{eq:RMC_Primakoff}). Those fitted spectra turned out to have an experimental value of $E^{end}_{\gamma}$ around 10 MeV smaller than the theoretical endpoint energy, which is calculated based on kinematics. Regarding this discrepancy, Eq. (\ref{eq:RMC_Primakoff}) was developed from the closure approximation, in which the excitation energy of a nucleus is averaged into a single energy. Theoretical attempts have been made to correct this spectrum assuming that the final nuclei are excited with dipole resonance or higher resonance modes within the nuclear collective model  \cite{CHRISTILLIN1981391, GMITRO1986}. However, because an uncertainty in the nuclear excitation model still remains, we utilize Eq. (\ref{eq:RMC_Primakoff}) with the kinematical endpoint energy (for example, 101.85 MeV for aluminum) for conservative estimation. 
\\
\indent The number of accepted background positrons from RMC $(N_{RMC})$ above the low end of the energy window for signal positrons ($E_{min}$) is given by
\begin{eqnarray}\label{eq:N_RMC}
N_{\textrm{RMC}} & \sim & N_{\mu^-\textrm{stop}} \times f_\textrm{cap} \times \textrm{Br(RMC)} \times P_{\gamma \rightarrow e^- + e^+} \nonumber \\ & & \times  P_{V \subset T} \times P_{E_{e^+}>E_{\textrm{min}}} \times \mathcal{E},
\end{eqnarray}
where $Br(\textrm{RMC})$ is the branching ratio of RMC whose photon energy is higher than $E_{min}$, $P_{\gamma \rightarrow e^- + e^+}$ is the probability of a pair production, $P_{V \subset T}$ is the probability that a pair production vertex is located inside the stopping target, and $P_{E_{e^+}>E_{min}}$ is the probability that a positron from the pair production has an energy higher than $E_{\min}$. 
Here, $P_{V \subset T}$ is included 
because the events that the vertex of pair production is located outside the stopping target can be avoided by using extrapolation of the positron tracks. There is another possibility that internal conversion could occur with an off-shell photon.  Since there have not been detailed studies on the energy spectrum of positrons emitted by the internal conversion, the amount of background was conservatively assumed to be the same as the on-shell RMC background throughout this paper. 
\\
\indent In the following subsections, simulation studies using Geant4 \cite{Geant4} with a muon-stopping target made of aluminum and target nucleus candidates in Table. \ref{tab:target_candidate} are shown, respectively.  
\subsection{\label{ssec:aluminum_estimation}Subcase: aluminum target}
For the sensitivity estimation, it is necessary to know the probability density functions (PDF) of positrons from both of the $\mu^--e^+$ conversion and RMC, including their normalization factors of PDF, i.e., $N_{\mu^-e^+}$ and $N_{RMC}$. The PDF of signal positrons was obtained by generating $10^4$ positrons with the energy of $E_{\mu^-e^+}$ in the aluminum muon-stopping target. The muon-stopping target was composed of 17 flat disks whose radius is 100 mm, thickness is 200 $\mu$m, and the spacing between disks is 50 mm, benchmarking the design of the COMET target \cite{COMET_CDR_2009}. The energy of positrons was measured after they exited the target to consider the energy loss in the target. The PDF of $f(E_{\mu^-e^+}-x)$, where $f(x)$ is the standard Landau distribution, was used to fit the signal positron distribution. The fitted PDF was normalized to $N_{\mu^-e^+}$, which is determined by the value of $Br(\mu^--e^+)$, while $N_{\mu^-\textrm{stop}}$ of $10^{18}$ and $\mathcal{E}$ of $10^{-2}$ were chosen to achieve the $\mu^--e^-$ conversion sensitivity of $\mathcal{O}(10^{-16})$ with the aluminum target, based on the specifications of the upcoming experiments \cite{COMET_CDR_2009,COMET_TDR_2015,Mu2e_TDR_2014}.
\\
\indent 
$N_{RMC}$ was obtained by generating $10^7$ photons with the RMC spectrum above 90.30 MeV $(E_{min})$ inside the aluminum stopping targets. Simulation results showed that $P_{\gamma \rightarrow e^-+e^+}$ is 0.97, $P_{V \subset T}$ is 0.0058, and $P_{E_{e^+}>E_{min}}$ is 0.018. $Br(\textrm{RMC})$ has a value of $6.22 \times 10^{-7}$ in an energy range from 90.30 MeV to 101.85 MeV according to the results of Ref. \cite{Measday2001}. By plugging these values into Eq. (\ref{eq:N_RMC}), $N_{RMC}$ is expected to be $3.8 \times 10^{5}$ without considering the internal conversions of the off-shell photons. For the PDF of RMC positrons, another simulation was done independently to obtain enough samples of positrons from RMC. The RMC photons of $2\times10^6$ with the same energy range were generated inside the large size of aluminum, in which almost half of the photons decay via pair productions. The PDF of positrons from the pair production was fitted to the power function of $A(E^{end}_{RMC}-x)^y$, where $A$ is the normalization constant, $x$ is the positron energy, and $y$ is the running parameter to be fitted. The PDF after fitting was forced to be zero above $E^{end}_{RMC}$, and convoluted by the Landau energy loss distribution $f(-x)$ of the signal positrons, assuming that the energy loss distributions of positrons from $\mu^--e^+$ and RMC would not be substantially different from each other. The convoluted PDF was normalized to $N_{RMC}$ afterwards.
\begin{figure}
\centering
    \includegraphics[width=0.49\textwidth]{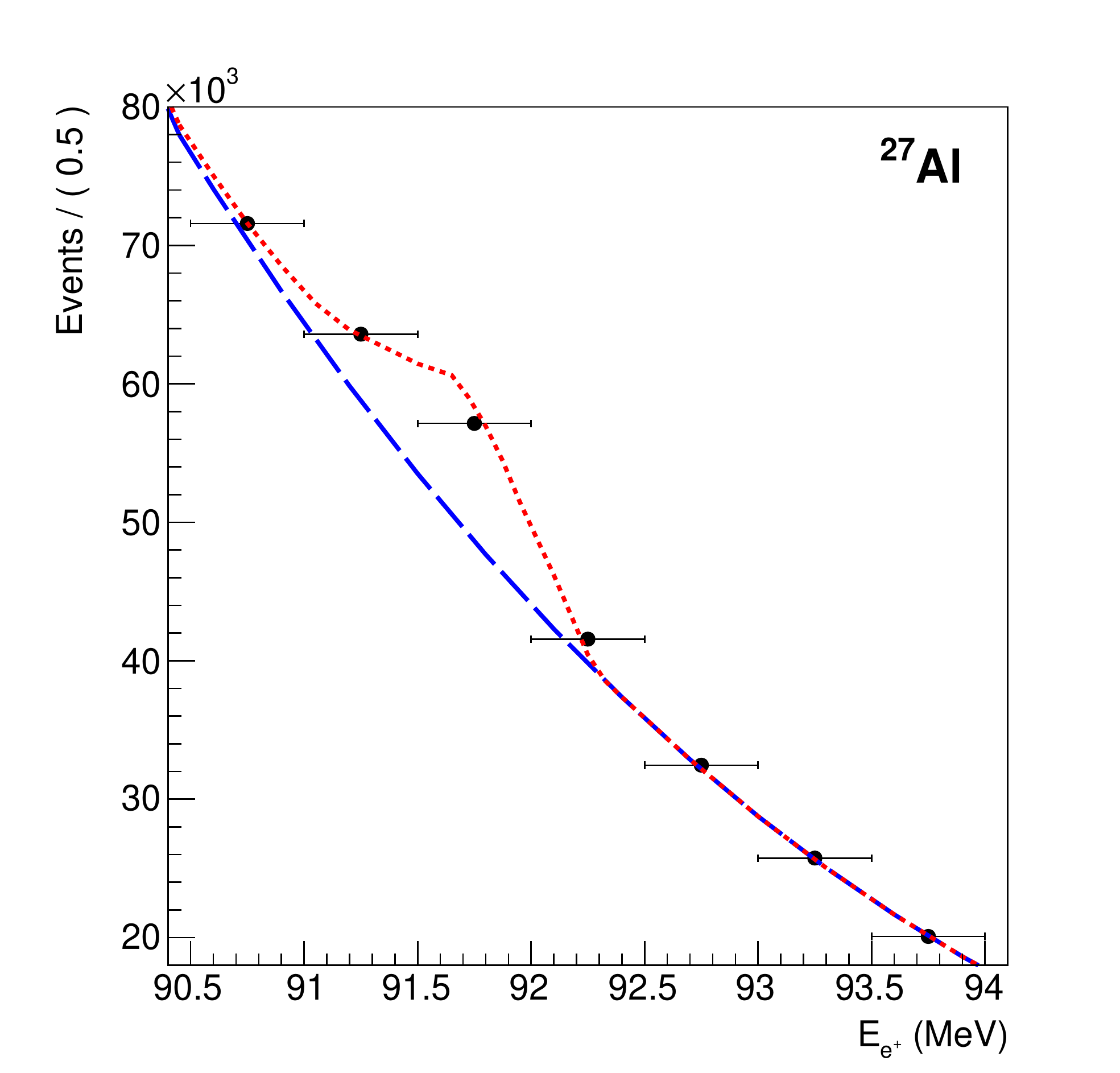}
\caption{Fitting result of the energy distribution of the $\mu^--e^+$ signal (short dashed red line) stacked on the on-shell RMC photon background (long dashed blue line) from $^{27}\textrm{Al}$ muon-stopping target when $Br(\mu^--e^+) = 1.7 \times 10^{-12}$ and $N_{\mu^{-} {\rm stop}} = 10^{18}$. Black dots are pseudo data of positrons generated by the background and signal composite model.}
\label{fig:Al_CompModel} 	
\end{figure}
\\
\indent For an illustrative purpose, Fig. \ref{fig:Al_CompModel} shows the estimation of the energy spectrum of the RMC background from on-shell photons and the $\mu^--e^+$ signal positron with a $Br(\mu^--e^+)$ of $1.7 \times 10^{-12}$, which is the current world-wide limit. The energy distributions of the signal and RMC were convoluted with a Gaussian detector response function with 200 keV standard deviation. To estimate the improvement of the sensitivity over the current limit, the statistical significance of the signal with a given $Br(\mu^--e^+)$ was examined using a maximum likelihood method. The systematical uncertainties were assumed to be negligible. With this assumption, $Br(\mu^--e^+)$, which has $3\sigma$ significance under the null hypothesis, was found to be $1.5 \times 10^{-13}$. When the internal conversions of RMC are included, $Br(\mu^--e^+)$ $<$ $2.5 \times 10^{-13}$ was found to have $3\sigma$ significance. These results imply that sensitivity improvement of more than a factor of ten, which is $Br(\mu^--e^+)$ $<$ $1.7 \times 10^{-13}$, may not be achieved with aluminum.

\subsection{\label{ssec:candidates_estimation}Target of the candidate nuclei}
\begin{figure}
\centering
    \includegraphics[width=0.49\textwidth]{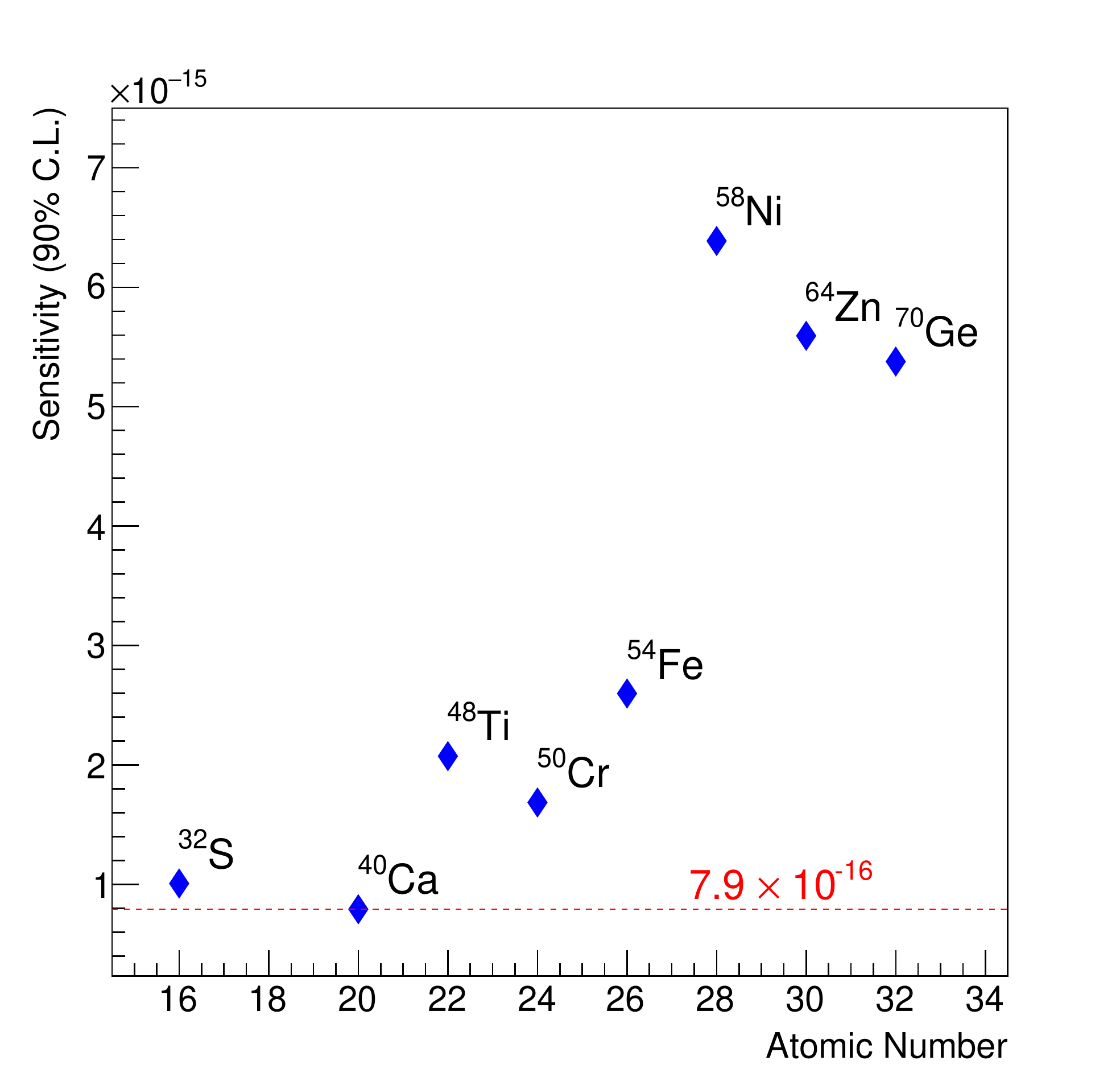}
\caption{Experimental sensitivities ($90\%$ C.L.) of the target nucleus candidates. Red dotted line and the number in red above the line indicate the sensitivity of $^{40}\textrm{Ca}$, which is the best among the target nucleus candidates.}
\label{fig:Sensitivities_Internal} 	
\end{figure}
\begin{figure*}
\centering
    \includegraphics[width=0.49\textwidth]{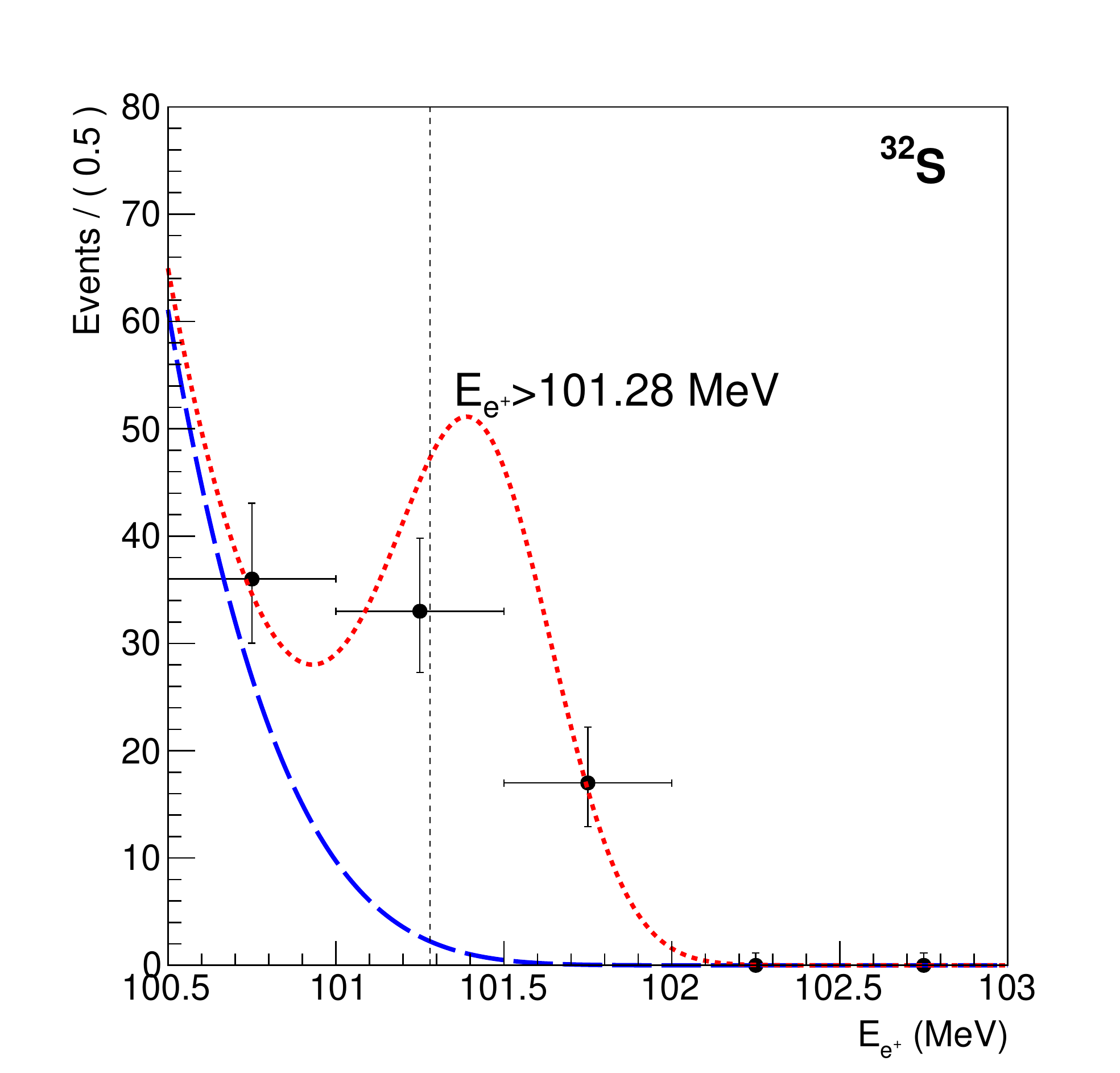}
    \includegraphics[width=0.49\textwidth]{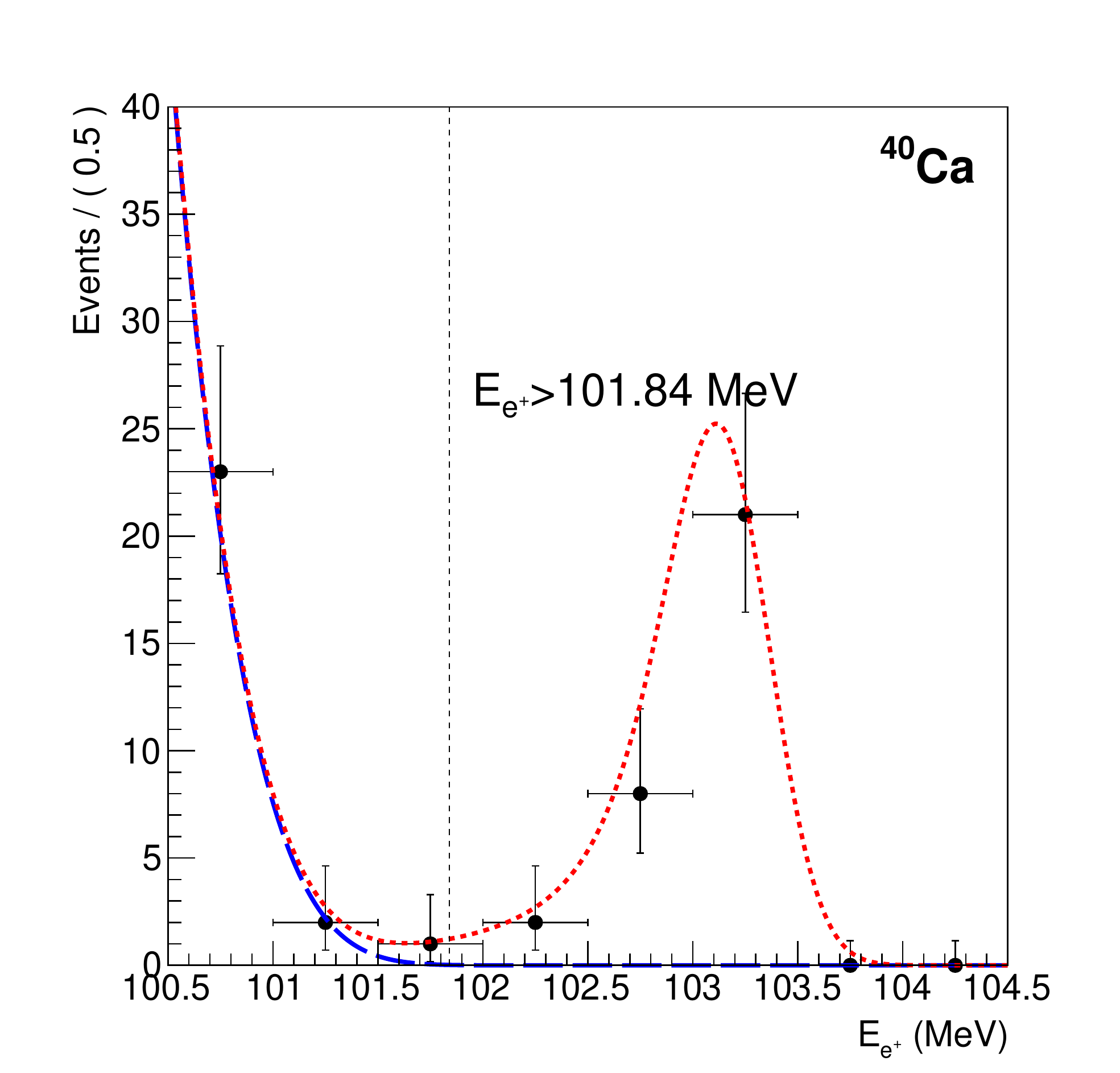}
    \includegraphics[width=0.49\textwidth]{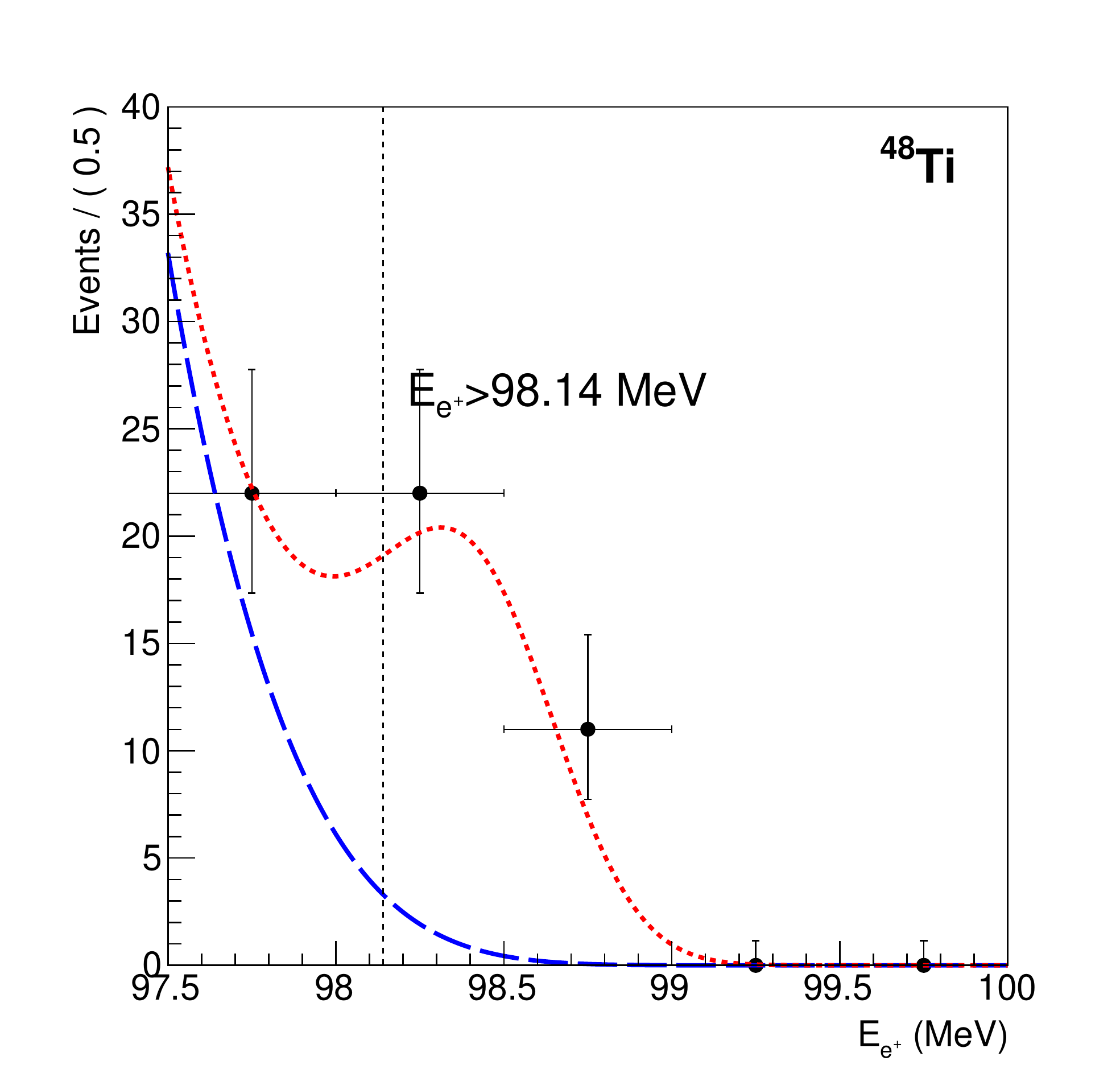}
    \includegraphics[width=0.49\textwidth]{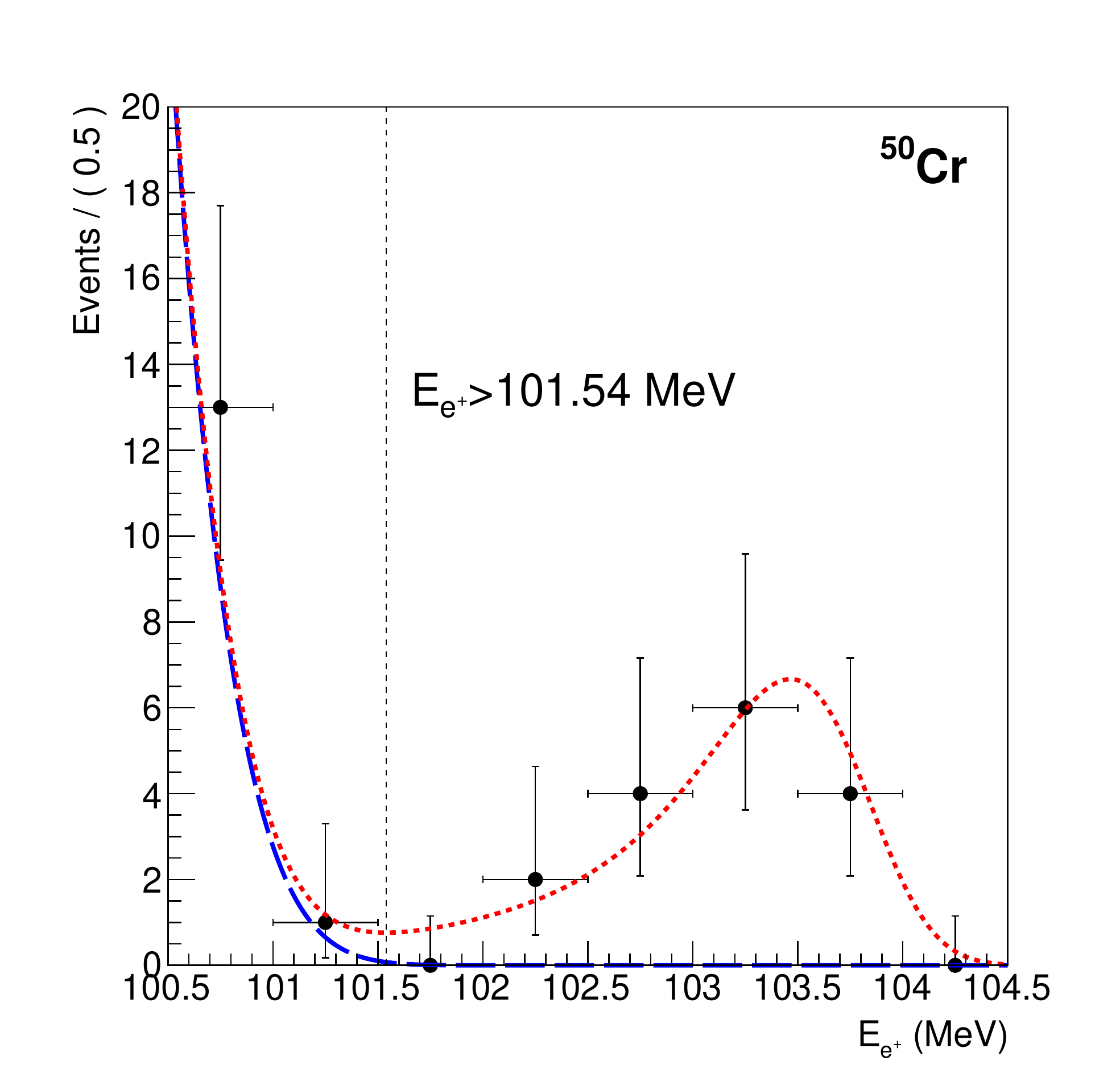}
\caption{Fitting result of the energy distributions of the $\mu^--e^+$ signal (short dashed red line) stacked on the RMC photon background (long dashed blue line) from $^{32}\textrm{S}$, $^{40}\textrm{Ca}$, $^{48}\textrm{Ti}$, and $^{50}\textrm{Cr}$ muon-stopping target when $Br(\mu^--e^+)=1.0\times 10^{-14}$ and $N_{\mu^{-} {\rm stop}} = 10^{18}$. The inequality beside the vertical black dotted line represents the signal energy window, and the line corresponds to its lower boundary. Black dots are pseudo data of positrons generated by the background and signal composite model.}
\label{fig:Ca_Internal} 
\end{figure*}
\begin{figure*}
\centering
    \includegraphics[width=0.49\textwidth]{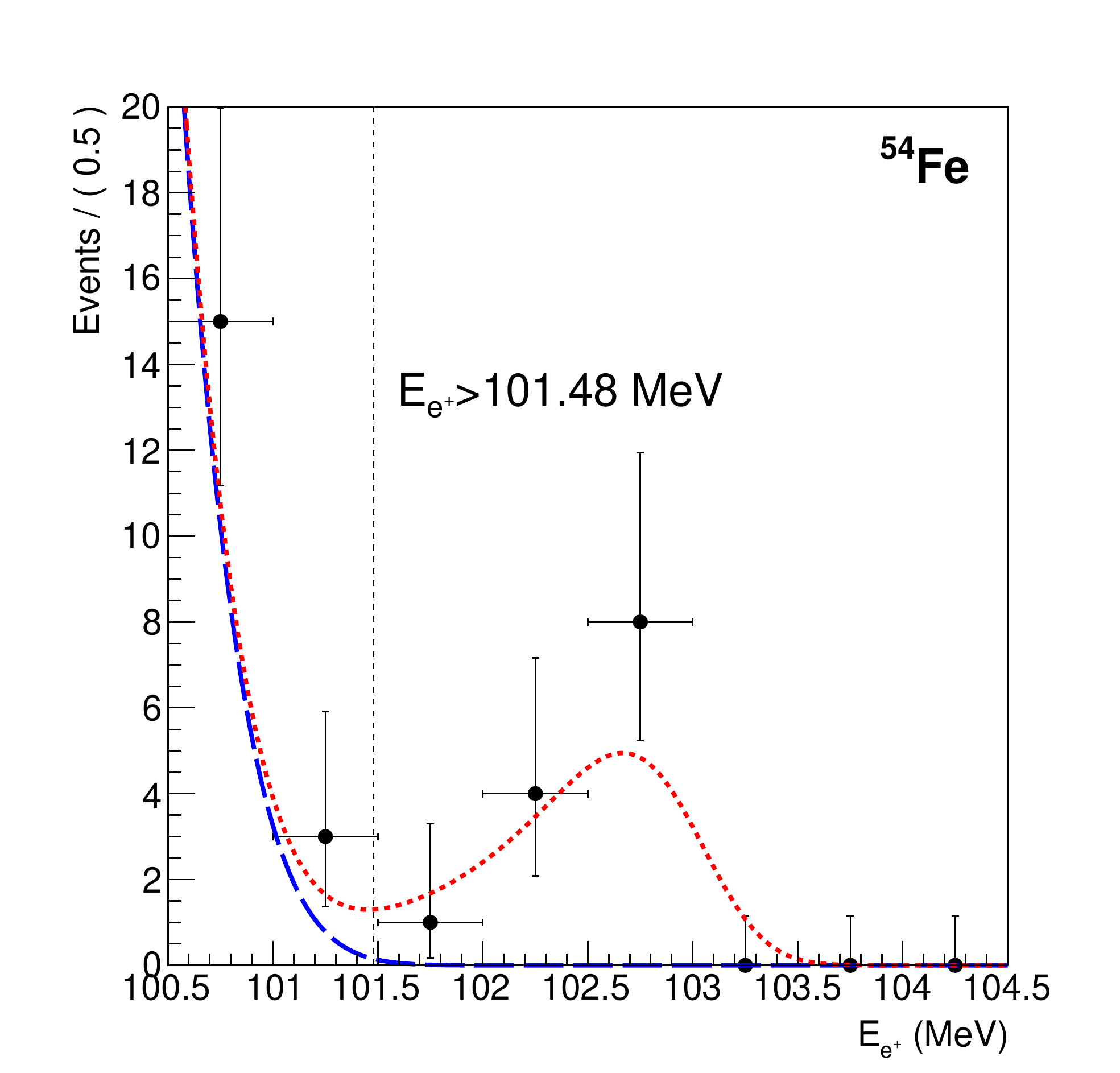}
    \includegraphics[width=0.49\textwidth]{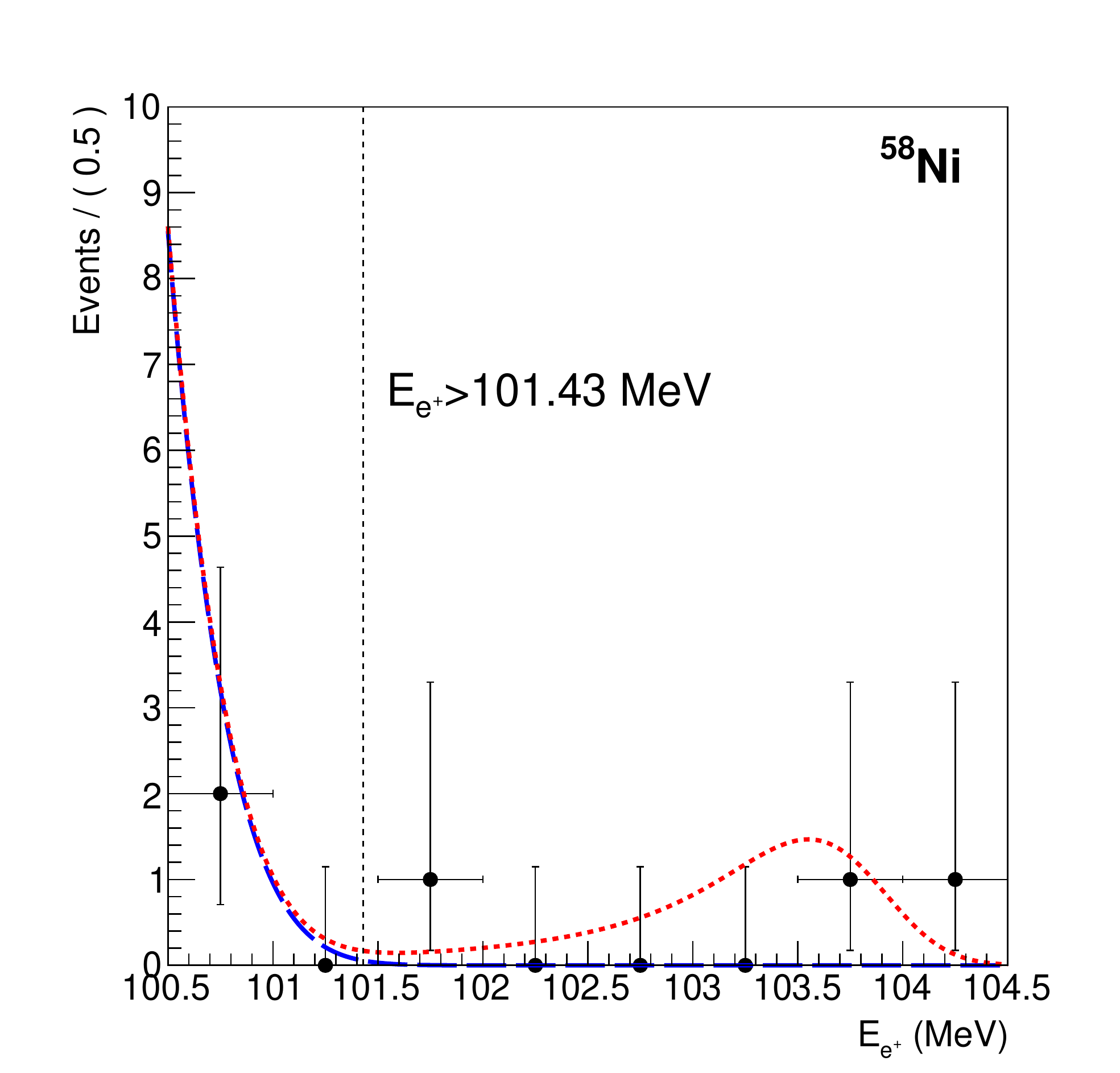}
    \includegraphics[width=0.49\textwidth]{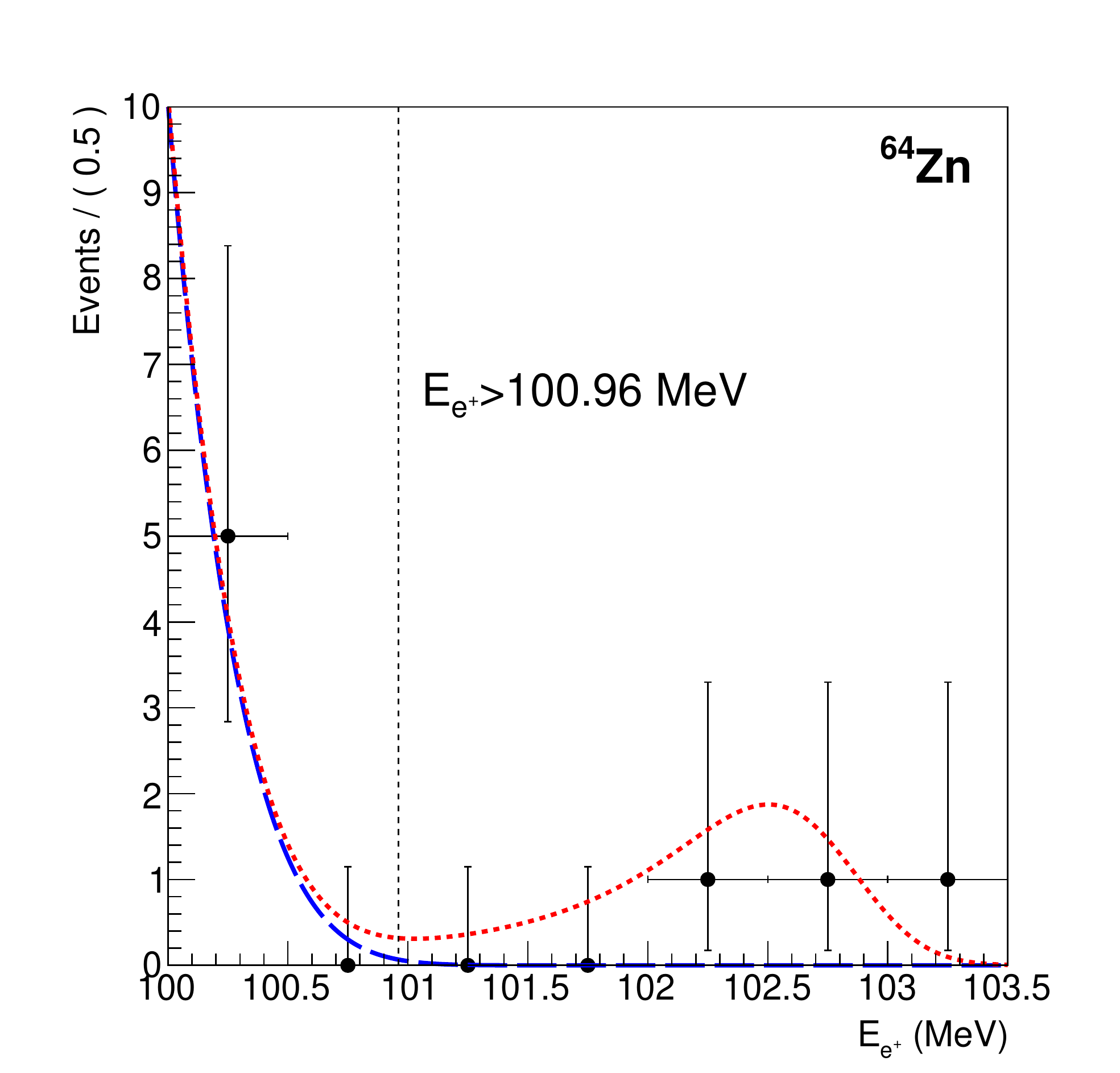}
    \includegraphics[width=0.49\textwidth]{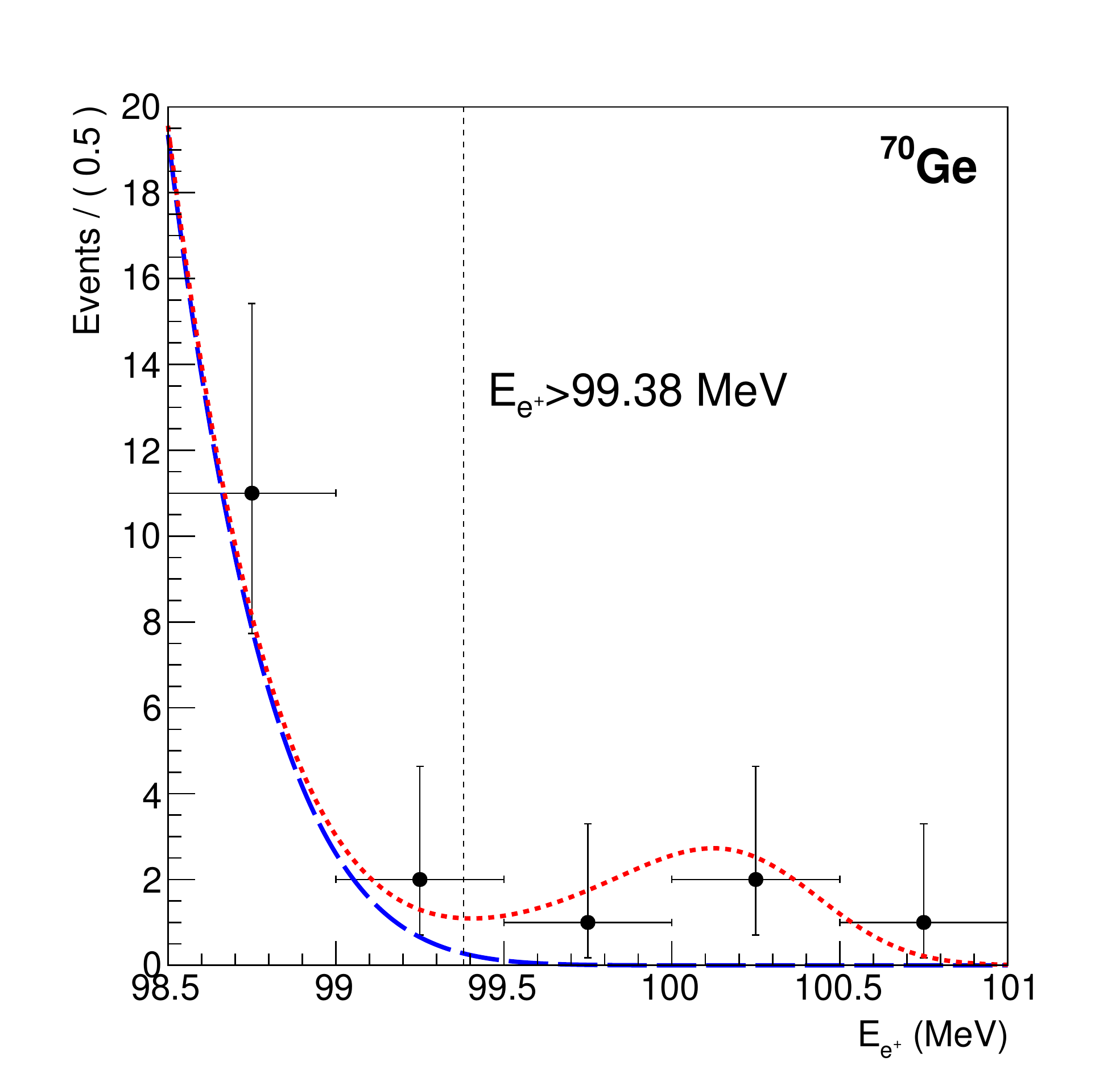}
\caption{Fitting result of the energy distributions of the $\mu^--e^+$ signal (short dashed red line) stacked on the RMC photon background (long dashed blue line) from $^{54}\textrm{Fe}$, $^{58}\textrm{Ni}$, $^{64}\textrm{Zn}$, and $^{70}\textrm{Ge}$ muon-stopping target when $Br(\mu^--e^+)=1.0\times 10^{-14}$ and $N_{\mu^{-} {\rm stop}} = 10^{18}$. The inequality beside the vertical black dotted line represents the signal energy window, and the line corresponds to its lower boundary. Black dots are pseudo data of positrons generated by the background and signal composite model.}
\label{fig:Fe_Internal} 
\end{figure*}

The same simulation and analysis for the target nucleus candidates in Table \ref{tab:target_candidate}
 were done with the corresponding $E_{\mu^-e^+}$ and $E^{end}_{RMC}$. Since the RMC branching ratios of $^{32}\textrm{S}$, $^{50}\textrm{Cr}$, $^{64}\textrm{Zn}$ and $^{70}\textrm{Ge}$ have not been measured, we used the known branching ratios of nuclei in Ref. \cite{Measday2001}, whose atomic number is closest to the relevant nucleus, i.e., $^{28}\textrm{Si}$ for $^{32}\textrm{S}$, Ti for $^{50}\textrm{Cr}$, and $^{58}\textrm{Ni}$ for $^{64}\textrm{Zn}$ and $^{70}\textrm{Ge}$. In the simulation, we assumed the target is made of a pure isotope candidate. $\mathcal{E}$ was normalized relative to that of aluminum $(10^{-2})$ by considering $A_T$ and 
the signal acceptance in the energy window, $A_E$, that the positron energy is in an acceptable energy range defined more strictly to count the number of events accurately. In other words, $\mathcal{E}\rightarrow \mathcal{E} \times 
A_E/A^{\textrm{Al}}_E \times A_T/A^{\textrm{Al}}_T$, where $A^{\textrm{Al}}_{E}$ and $A^{\textrm{Al}}_{T}$ are $A_{E}$ and $A_{T}$ of aluminum, respectively. It should also be noted that the values of $A_E$ for Eq. (\ref{eq:N_SIG}) and Eq. (\ref{eq:N_RMC}) are different from each other because of the difference in the two PDFs. 
\\
\indent 
Since the numbers of the RMC background events for these candidate targets are much less than the aluminum case, the experimental sensitivities were estimated as the upper limit of a $90\%$ confidence level for a direct comparison with the current experimental upper limit of $1.7\times10^{-12}$ ($90\%$ C.L.). The number of observed events $(n)$ 
follows a Poisson distribution given by $p(n|s)=(s+b)^n e^{-(s+b)}/n!$, where $s$ and $b$ are the expected numbers of events of signal and background, which are equivalent to $N_{\mu^-e^+}$ and $N_{RMC}$, respectively. The experimental sensitivity with a confidence level $(\alpha)$, defined as the average upper limit of repeated experiments with no true signal, is given by the following equation: 
\begin{equation}
\frac{\int^{s_{\textrm{up}}}_{0}p(n=b|s)ds}{\int^{\infty}_{0}p(n=b|s)ds}=\alpha,
\end{equation}
where $s_{up}$ is the upper limit of the expected number of the signal events, which can be converted into the upper limit of the branching ratio from Eq. (\ref{eq:N_SIG}). The upper limit for each target nucleus was optimized by tuning the signal energy window ($A_{E}$) because both $N_{\mu^-e^+}$ and $N_{RMC}$ are dependent on $A_{E}$. Figure \ref{fig:Sensitivities_Internal} shows the experimental sensitivity ($90\%$ C.L) of each target nucleus, when the internal conversion of the RMC background is included as well. As a result, $^{40}\textrm{Ca}$ showed the best experimental sensitivity, $7.9\times 10^{-16}$, among the candidates investigated, followed by $^{32}\textrm{S}$ with the sensitivity of $1.0 \times 10^{-15}$. The energy distributions of positrons from each nucleus are shown in Fig. \ref{fig:Ca_Internal} and Fig. \ref{fig:Fe_Internal}, in which the branching ratio of the $\mu^--e^+$ conversion is set to $1.0 \times 10^{-14}$, and $N_{\mu^-\textrm{stop}}$ is set to $10^{18}$ as the aluminum case. 

\section{Summary}\label{sec:summary}
A profound understanding of leptons is important because the fundamental conservation laws of leptons within the SM are easily violated in most of the theoretical models beyond the SM. 
Among them, LNV processes are important tools to reveal the mechanism of the neutrino mass generation. Investigation of the LNV processes mostly has been conducted through $0\nu\beta\beta$ decay experiments, but the experimental search for the $\mu^--e^+$ conversion can also be carried out as a complementary channel to the $0\nu\beta\beta$ decay. Since a great leap of the sensitivity of the $\mu^--e^+$ conversion is expected with the future CLFV experiments, it is essential to make a full exploration of the current experimental scheme. \\
\indent
For this purpose, we introduced a new requirement of the target nucleus mass of $M(A,Z)$ satisfying $M(A,Z-2)<M(A,Z-1)$ to suppress the backgrounds from RMC. Several appropriate target candidates of even-even nuclei were found to meet the criteria. We estimated the experimental sensitivities of such target nuclei candidates in a general experimental set-up. In conclusion, calcium ($^{40}$Ca) and sulfur ($^{32}$S) have the best experimental sensitivities about $\mathcal{O}(10^{-16})$ in the $\mu^--e^+$ conversion detection, which results in a four orders of magnitude of improvement compared to the current upper limit. Another advantage of these two materials is that they will also have better sensitivities in the $\mu^--e^-$ conversion measurement due to their relatively high timing efficiencies. It should be noted that the actual sensitivity would be different in the real experiment because some factors such as systematical uncertainties are not considered in this paper. However, this result can be a useful standard in the selection of the muon-stopping target material in future experiments.
\begin{acknowledgments}
This work of B.Y. and M.L. was supported by the Institute for Basic Science of Republic of Korea (project number: IBS-R017-D1-2017-a00). This work of Y.K. was supported in part by the Japan Society of the Promotion of Science (JSPS) KAKENHI Grant No. 25000004. We would like to thank the National Institute of Nuclear and Particle Physics (IN2P3) for providing the computing resources.
\end{acknowledgments}

\bibliography{references}

\end{document}